\documentclass[prl,showpacs, nobibnotes]{revtex4}
\usepackage{color,graphicx,epsf}
\def\red#1 {{\em\color{red}#1 }}
\begin{document}

\title{Magnetism in BN Nanotubes Induced by Carbon Substitution}

\author{R. Wu, G. Peng, L. Liu, Y. P. Feng}
\email{phyfyp@nus.edu.sg}
\affiliation{Department of Physics, National University of Singapore, 2 Science Drive 3, Singapore 117542}

\begin{abstract}
We performed {\em ab initio} calculation on the pristine and carbon-doped (5,5) and (9,0) 
BN nanotubes. It was found that Carbon substitution for either boron or nitrogen in BN 
nanotubes can induce spontaneous magnetization. 
Calculations based on density functional theory with the local-spin-density-approximation  
on the electronic band structure revealed a spin polarized, dispersionless band near 
the Fermi energy. The magnetization can be attributed to the carbon $2p$ electron. 
Compared to other theoretical models of light-element or metal-free magnetic 
materials, the Carbon-doped BN nanotubes are more experimentally accessible and can be
potentially useful.
\end{abstract}
\maketitle

The discovery of room-temperature weak ferromagnetism in all-carbon system consisting of 
polymerized C$_{60}$ \cite{Makarova} stimulated wide interest in 
magnetism of all-carbon materials. 
One year later after the above discovery, Esquinazi {\em et al.}\cite{Esquinazi} detected a ferromagnetic signal 
from an oriented graphite which behaved quite differently from known magnetic impurities, 
suggesting an intrinsic origin of magnetism in graphite. 
In order to understand the unexpected magnetism in all-carbon system, 
some density functional theory (DFT) calculations have been carried out. 
Lehtinen {\em et al.}\cite{Lehtinen} performed {\em ab initio} local-spin-density approximation (LSDA) 
calculations to study the properties of a carbon adatom on a graphite sheet and 
found that this defect has a magnetic moment of up to 0.5 $\mu_B$. 
This was important in understanding the magnetism observed experimentally in graphite sheet.
More recently, Ma {\em et al.}\cite{Ma} studied the magnetic properties of vacancies in graphite 
and carbon nanotubes. 
For graphite, the vacancy is spin-polarized with a magnetic moment of 1.0 $\mu_B$. 
The vacancy in carbon nanotubes can also induce magnetism, depending on the chirality of the 
nanotubes and the structural configuration with respective to the tube axis. 
These calculations provided some understanding of the observed magnetism in pure carbon systems.

The weak ferromagnetism in pure carbon system also stimulated interest in searching for 
light-elements or metal-free magnetic materials in view of their potential applications as high 
temperature magnets since metal magnets lose their ferromagnetism at high temperatures. 
One of such metal-free magnets was nanographite or graphite-ribbon. 
Fujita {\em et al.} \cite{Fujita} performed tight binding 
band structure calculations on graphite ribbons with armchair and zigzag edges, respectively. 
Ribbons with zigzag edges show a sharp peak in density of states at Fermi level, indicating a 
possibility of spontaneous magnetization. 
Mono-hydrogenation such as zigzag-edged graphite ribbon could create a ferromagnetic spin 
structure on their edges. 
Based on results of first principle pseudopotential calculation, Kusakabe and 
Maruyama\cite{Kusakabe} further predicted that the zigzag edged graphite 
ribbon could also have finite magnetization at edges by hydrogenating each carbon with two 
hydrogen atoms while hydrogenating each carbon with a single hydrogen atom at the other edge. 
However, such a hydrogenation of carbon edges is only of interest for theorists. 
Another recently proposed model by Choi and coworkers\cite{Choi} is the heterostructured 
C-BN nanotube. 
They calculated the electronic structure of the (9,0) C$_1$(BN)$_1$ and C$_2$(BN)$_2$ nanotubes 
using density functional theory and found the occurrence of magnetism at the zigzag boundary 
connecting carbon and boron nitride segments of tubes.
However, in view of the growth conditions of carbon and boron nitride nanotubes, fabrication of 
the heterostructured C-BN nanotubes may be impractical.

In this letter, we report results of our DFT-LSDA studies on the electronic 
structures of the pristine and carbon-doped (5,5) and (9,0) BN nanotubes. 
We find that the carbon substitution for either boron or nitrogen atom induces 
magnetization of the doped system, with a total magnetic moment of 1 $\mu_B$. 
Compared to previous models of metal-free materials, our proposed C-doped BN nanotubes 
are experimentally accessible as far as the structural configuration is concerned.

Our calculation is base on the density functional theory and the local spin density functional formulism\cite{Perdew}. 
The calculation was done using the SIESTA code\cite{Ordejon}. 
The valence electrons are described by linear combination of numerical 
atomic-orbital basis set and the atomic core by norm-conserving pseudopotentials. 
The pseudopotentials generated using the Troullier and Martins scheme\cite{Troullier} 
are used to describe the interaction of valence electrons with atomic core and 
their nonlocal components are expressed in the fully separable form of Kleiman and 
Bylander \cite{Kleiman,Bylander}. 
The generalized gradient approximation correction in the form of Perdew, Burke and 
Ernzerhof\cite{PBE} is adopted for the exchange-correlation potential. 
The double-zeta plus polarization atomic orbital basis set is employed in the calculation. 
The Hamiltonian matrix elements are calculated by charge density projection on 
a real space grid with an equivalent plane wave cutoff energy of 70 Ry. 
Periodical boundary condition is employed along the nanotube axis, and 
a vacuum region of at least 10 \AA\ is assumed between the nanotube and its images,
to avoid tube-tube interaction.
Special k-mesh is chosen along the tube axis according to the Monkhorst-Pack scheme\cite{MP}. 
The conjugate gradient algorithm \cite{Press} was adopted to fully relax the structure
of the nanotube until the maximum force on a single atom is within $ 0.02$ eV/\AA.
For the calculation of carbon doped (5,5) nanotube, we use a supercell which consists of 
$1\times 1\times 2$ primitive cells and contains 80 atoms to avoid possible coupling 
between carbon images.

The calculated band structures of pristine (5,5) and (9,0) BN nanotubes are shown in Fig.~1. 
The band gaps of 4.47 eV for (5,5) and 3.81 eV for (9,0) nanotubes agree well 
with those of {\em ab initio} calculation by Xiang {\em et al.}\cite{Xiang} 
The (5,5) BN nanotube has an indirect band gap while the (9,0) has a direct band gap. 
The calculation also reproduces the free-electron-like lowest conduct bands as found 
by Blas\'e {\em et al.} \cite{Blase} 
These results confirm the validity of the application of atomic orbital basis set. 
We also performed spin-polarized calculation on these two pristine nanotubes and no 
spontaneous magnetization was found.

However, when either boron or nitrogen is substituted by carbon, a spontaneous 
magnetization is induced. The band structure of the (5,5) BN nanotube with a boron atom 
substituted by carbon atom is shown in Fig.~2(a). 
We can see that the carbon substitution causes a lift up of the Fermi level. 
All bands with energies lower than -6.0 eV are spin degenerate and fully occupied,
and thus do not contribute to spin-polarization. 
However, close to the Fermi level, a very flat band, almost dispersionless in the whole first 
Brillouin zone, is split. 
Only the spin-up branch is occupied and the spin-down branch is left unfilled, 
leading to a strong spontaneous magnetization in the (5,5) nanotube. 
The splitting between the occupied and unfilled branch is as large as 1.5 eV, 
The flatness of the spin-polarized band indicates that the corresponding electron state is 
heavily localized, suggesting that the magnetic moment is localized at the substitution site, 
possibly from the carbon 2p electron. 
We also investigated single carbon substitution for nitrogen on the (5,5) 
nanotube and found that it also produces a spin polarized band around the Fermi level. 
However, the Fermi level is pushed down. The net magnetic moment is about 1 $\mu_B$ in both substitutions.

The band structure of the (9,0) nanotube with a nitrogen atom substituted by carbon atom
is shown in Fig.~2(b). In this case, the Fermi level is pushed down, similar to the case of
carbon substitution for nitrogen in the (5,5) nanotube. A spin-polarized band occurs around 
the Fermi level with the spin-up branch occupied and the spin-down branch unfilled, resulting 
in a total magnetic moment of 1 $\mu_B$. 
Similar calculation was carried out for the case of carbon substitution for boron atom 
and it was found that the Fermi level increases and the band below the Fermi level is partially
occupied and spin-polarized, leading to a net magnetic moment of 1 $\mu_B$. 

The spin polarization can also be clearly seen from the spin density of states (DOS),
as shown in Fig.~3. 
For carbon substituted (5,5) nanotube (Fig.3~(a)), a separate occupied majority spin peak 
emerges at the energy of $-4.0$ eV. 
Below this peak, the majority and minority spin density of states are essentially 
identical. 
For the carbon substituted (9,0) nanotube (Fig.3~(b)), we also can see that near the 
Fermi level, the majority spin DOS exceeds that of the minority. 
However, the highest occupied spin-up branch merges with the lower valence bands in 
this case.

Boron nitride has important advantages over carbon nanotube. 
It is far more resistant to oxidation than carbon and therefore more suitable 
for high-temperature applications in which carbon nanostructures would burn.
Unlike carbon nanotubes, BN nanotubes are insulators, with predictable electronic properties 
that are independent of their chiralities. In addition,
magnetic nanostructures are of scientifically interesting and technologically important, 
with many present and future applications in permanent magnetism, magnetic
recording and spintronics. The carbon substituted BN nanotubes, with conduction 
electrons that are 100\% spin polarized due to the gap at the Fermi level in one
spin channel, and a finite density of states for the other spin channel,
can be an ideal half metallic material and can be useful for spintronic applications, 
such as tunneling magnetoresistance and giant magnetoresistance elements.
Upon geometry optimization, all doped nanotubes were spontaneously spin polarized and 
no visible deformation or buckling in their structures can be observed. 
Contrast to the models proposed by Kusakabe \cite {Kusakabe} and Choi \cite {Choi} 
in which the spin polarization only occurs for certain particular structural configurations, 
the spin polarization in BN nanotubes is induced by substitution with arbitrary atomic 
configuration, indicating easy accessibility of experimental synthesis.

In conclusion, we have performed first principle pseudopotential calculation with 
double-zeta plus polarization atomic-orbital basis set to study the effects of carbon 
substitution on BN nanotubes. Our results show that single carbon atom substitution 
for any atom results in a polarized flat band and a sharp peak 
in the majority density of state below the Fermi level. 
The spontaneous spin-polarization is independent of site of substitution and 
chirality of the nanotubes.  
Compared to other metal-free magnets previously proposed, the carbon induced magnetization
in BN nanotubes is experimentally accessible and the system can be potentially very useful.

\clearpage
\begin{figure}
\includegraphics{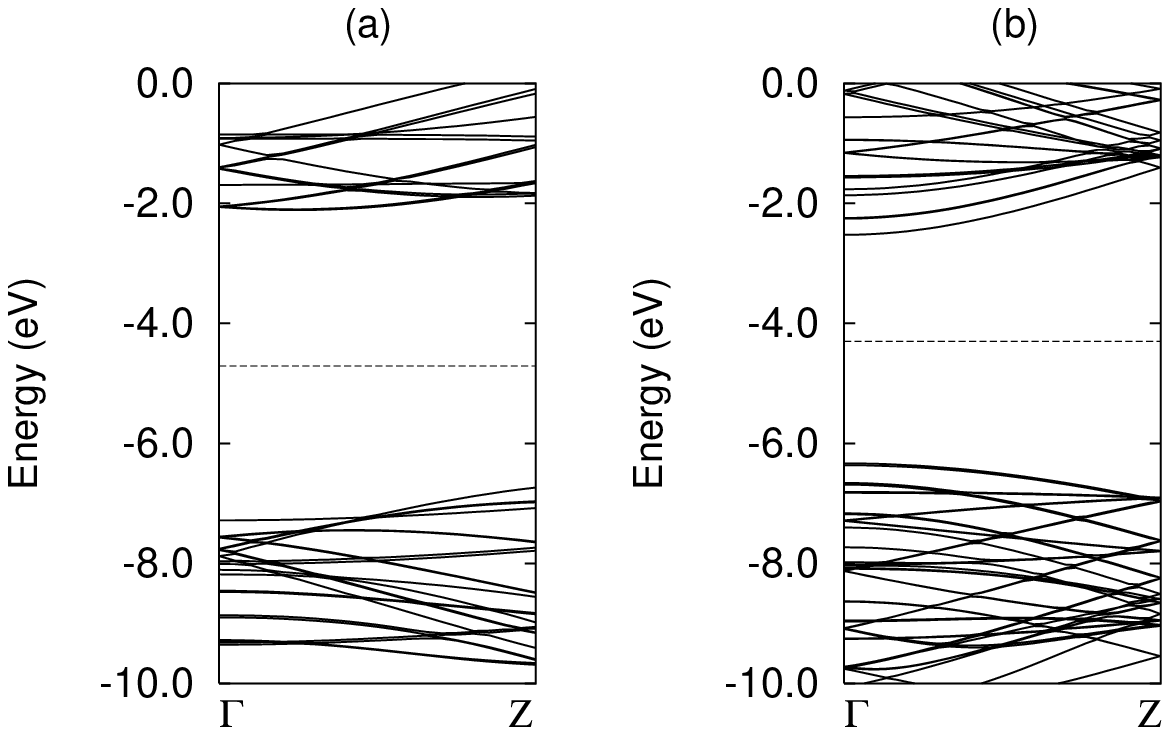}
\caption{The band structure of the pristine (5,5) (a) and (9,0) (b) BN nanotubes.}
\end{figure}

\clearpage 
\begin{figure*}
\begin{center}
\epsfbox{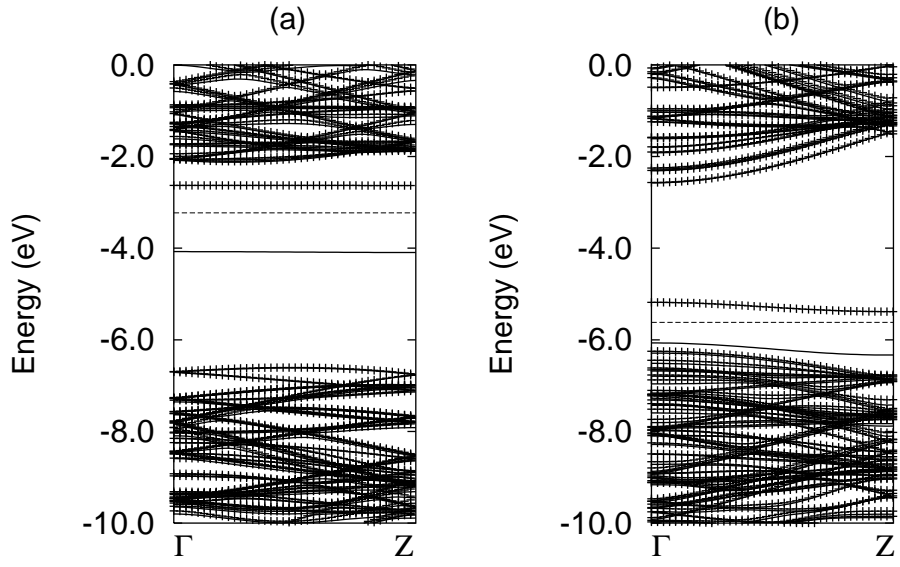}
\end{center}
\caption{(a) The band structure of (5,5) nanotube with a boron atom substituted by carbon. (b) 
The band structure of (9,0) nanotube with a nitrogen substituted by carbon. Solid lines and 
$+$ represent the bands for spin-up and spin-down electrons respectively. The Fermi level is 
denoted by the dotted line. }
\end{figure*}

\clearpage
\begin{figure*}
\includegraphics{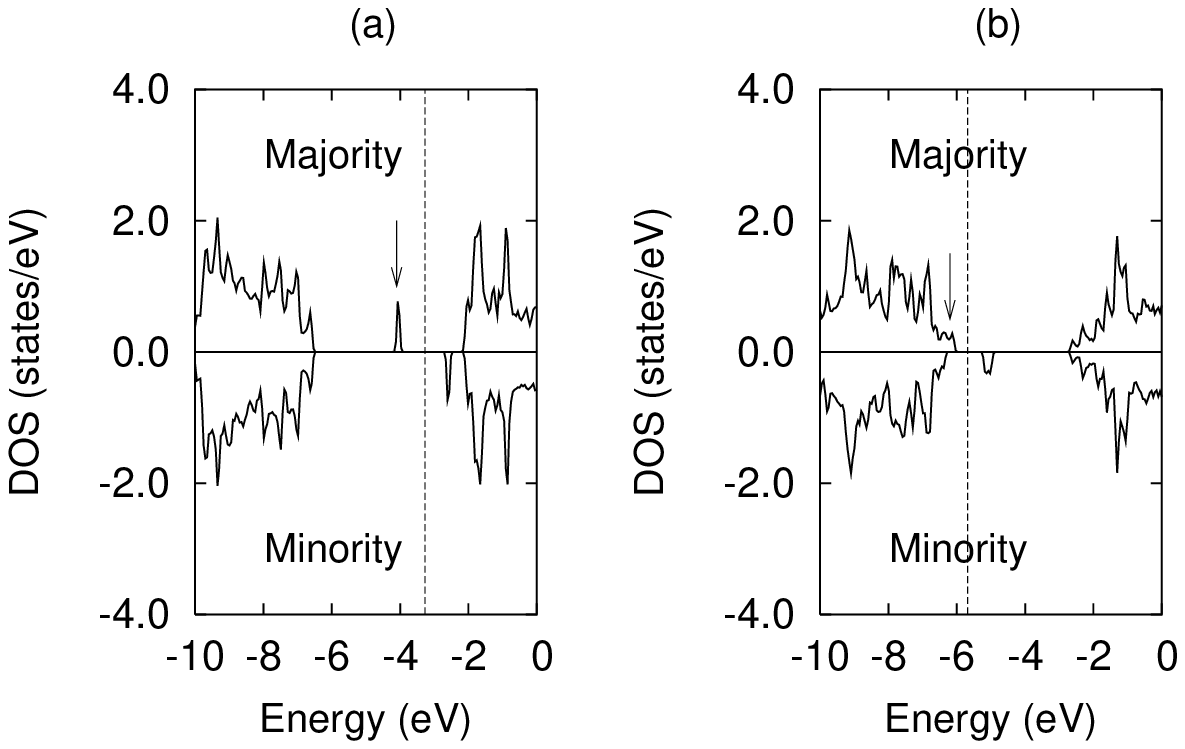}
\caption{Majority and minority spin densities of states of the (5,5) (a) and (9,0) 
BN nanotubes.}
\end{figure*}


\begin{thebibliography}{}

\bibitem{Makarova}
T. L. Makarova, B. Sundqvist, R. H\"ohne, P. Esquinazi, Nature, {\bf 413}, 716 (2001).
\bibitem{Esquinazi}
P. Esquinazi, A. Setzer and R. Hohne, Phys. Rev. B {\bf 66}, 024429 (2002).
\bibitem{Lehtinen}
P. O. Lehtinen, A. S. Foster and A. Ayuela, Phys. Rev. Lett {\bf 91}, 017202 (2004).
\bibitem{Ma}
Yuchen Ma, P. O. Lehtinen and A. S. Foster, New J. Phys {\bf 6}, 68 (2004).
\bibitem{Fujita}
M. Fujita, K. Wakabayashi, K. Nakata and K. Kusakabe, J. Phys. Soc. Jap {\bf 65}, 1920 (1996).
\bibitem{Kusakabe}
K. Kusakabe and M. Maruyama, Phys. Rev. B {\bf 67}, 092406 (2003).
\bibitem{Choi}
J. Choi, Y. H. Kim, K. J. Chang and D. Tomenek, Phys. Rev. B {\bf 67}, 125421 (2003).
\bibitem{Perdew}
J. P. Perdew and A. Zunger, Phys. Rev. B {\bf 23}, 5048 (1981).
\bibitem{Ordejon}
P. Ordejon {\em et al.}, Phys. Rev. B {\bf 53}, R10441 (1996).
\bibitem{Troullier}
N. Troullier and J. L. Martins, Phys. Rev. B {\bf 43}, 1993 (1991).
\bibitem{Kleiman}
L. Kleiman and D. M. Bylander, Phys. Rev. Lett. {\bf 48}, 1425 (1982).
\bibitem{Bylander}
D. M. Bylander and L. Kleiman, Phys. Rev. B {\bf 41}, 907 (1990).
\bibitem{PBE}
J. Perdew, K. Burke and M. Ernzerhof, Phys. Rev. Lett. {\bf 77}, 3865 (1996); {\bf 78}, 1396 (1997).
\bibitem{MP}
H. J. Monkhorst and J. D. Pack, Phys. Rev. B {\bf 13}, 5188 (1976).
\bibitem{Press}
W. H. Press, B. P. Flannery, S. A. Teukolsky and W. T. Vetterling, New Numerical 
Recipes (Cambridge University Press, New York, 1986).
\bibitem{Xiang}
H. J. Xiang, Jinlong Yang, J. G. Hou and Qingshi Zhu, Phys. Rev. B {\bf 68}, 035427 (2003).
\bibitem{Blase}
X. Blas\'e, A. Rubio, S. G. Louie and M. L. Cohen, Europhys. Lett. {\bf 28}, 335 (1994).
\end{thebibliography}
\end{document}